\documentclass[cits,a4paper]{PoS}

\usepackage{amsmath,amssymb}
\usepackage{subfigure}
\usepackage{graphicx}

\newcommand{\tr}{\operatorname{Tr}}
\newcommand{\re}{\operatorname{Re}}

\graphicspath{{plots/}}

\title{Lattice QCD with two light Wilson quarks and maximally twisted mass}

\ShortTitle{Lattice QCD with two light Wilson quarks and maximally twisted mass}

\author{\speaker{Carsten Urbach}%
  \thanks{Current address: Humboldt-Universit{\"a}t zu Berlin, Institut
  f{\"u}r Physik, Newtonstr. 15, 12489 Berlin, Germany}\\
  Theoretical Physics Division,\\
  Department of Mathematical Sciences,\\
  University of Liverpool,\\
  Liverpool L69 3BX, UK\\
  E-mail: \email{Carsten.Urbach@physik.hu-berlin.de}}

\author{for the European Twisted Mass Collaboration (ETMC)}

\abstract{We summarise status and recent results of the European Twisted Mass
  collaboration (ETMC). The collaboration has been generating
  gauge configurations for three different values of the lattice
  spacing $a\lesssim0.1\ \mathrm{fm}$ and values of the charged pseudo
  scalar mass as low as $300\ \mathrm{MeV}$ with two flavours of
  maximally twisted mass quarks.
  We provide evidence that $\mathcal{O}(a)$ improvement works very
  well with maximally twisted mass fermions and that also higher order
  lattice artifacts appear to be small. The currently only quantity in
  the light meson and baryon sector where cut-off effects are visible
  is the neutral pseudo scalar mass and we present an attempt to
  understand this from a theoretical point of view.

  We describe finite size effects and quark mass dependence of the
  mass and decay constant of the (charged) pseudo scalar meson
  with chiral perturbation theory formulae and our current estimate
  for the low energy constants $\bar{\ell}_{3,4}$ is
  $\bar{\ell}_3=3.44(8)(35)$ and $\bar{\ell}_4=4.61(4)(11)$. Results
  for the average up-down, the strange and the charm quark mass and
  the chiral condensate are also presented.}

\FullConference{The XXV International Symposium on Lattice Field Theory\\
		 July 30 - August 4 2007\\
		 Regensburg, Germany}

\begin{document}

\section{Introduction}

Whenever results obtained from lattice QCD simulations are to be
confronted with experimental results it is important to have a sound
control of systematic uncertainties emerging in lattice QCD. The most
prominent of those are discretisation errors, finite size effects
(FSE) and uncertainties arising from the unphysically large mass values
usually simulated. The main reason for their prominence is the fact
that lattice QCD simulations become increasingly computer time
demanding when a) the lattice spacing is reduced b) the quark masses
are reduced towards the physical point and c) the volume is
increased.

The control of these systematic uncertainties requires simulations
with an $\mathcal{O}(a)$ improved lattice formulation at
sufficiently small values of the lattice spacing $a$, say $a\lesssim0.1\
\mathrm{fm}$ where $\mathcal{O}(a^2)$ lattice artifacts are small.
Physical volumes should be large enough, say with spatial box size $L$
larger than $2\ \mathrm{fm}$ and $m_\mathrm{PS}\cdot L\gtrsim 3$ 
($m_\mathrm{PS}$ is the mass of the lightest pseudo scalar
particle). And, in order to be able to utilise chiral perturbation
theory ($\chi$PT) to bridge between simulated quark masses 
and the physical point, simulations with a range of masses are needed,
with the smallest value of $m_\mathrm{PS}\lesssim 300\
\mathrm{MeV}$. It goes without saying that the aforementioned bounds
are only estimates and need to be checked carefully in actual
simulations.

Due to recent algorithmic improvements
\cite{Hasenbusch:2001ne,Peardon:2002wb,AliKhan:2003br,Luscher:2004rx,Urbach:2005ji,Clark:2006fx}
it became possible to meet all these requirements using Wilson's
original formulation of lattice QCD. It has the advantage of being
conceptually clear and simple. And, $\mathcal{O}(a)$ improvement can be
implemented in several ways, one of which is using so called Wilson
twisted mass fermions \cite{Frezzotti:2000nk} at maximal twist. As was
shown in Ref.\cite{Frezzotti:2003ni}, in maximally twisted mass
lattice QCD (Mtm-LQCD) physical observables can be
obtained $\mathcal{O}(a)$ improved by tuning a single parameter only.
In particular, no operator specific improvement coefficients
need to be computed. This theoretical expectation could be verified in
the quenched approximation to work very well
\cite{Jansen:2003ir,Jansen:2005gf,Jansen:2005kk,Abdel-Rehim:2005gz} 
(for a recent review see Ref.~\cite{Shindler:2007vp}.)

Based on these successes in the quenched approximation the European
Twisted Mass (ETM) collaboration decided to start a large scale
simulation project using two flavours of mass degenerate quarks with the
lattice formulation of Mtm-LQCD. First accounts of this effort are published in
Refs.~\cite{Jansen:2006rf,Shindler:2006tm,Boucaud:2007uk} indicating
that $\mathcal{O}(a)$ improvement works very well when sea quark effects
are taken into account in the simulations. This proceeding
contribution aims to summarise the progress and current status of the
two flavour project of the ETM collaboration.

\section{Gauge and Fermionic Action}
\label{sec:action}

In the gauge sector we employ the so-called tree-level Symanzik
improved gauge action (tlSym) \cite{Weisz:1982zw}, viz.
\[
S_g = \frac{\beta}{3}\sum_x\left(  b_0\sum_{\substack{
    \mu,\nu=1\\1\leq\mu<\nu}}^4\{1-\re\tr(U^{1\times1}_{x,\mu,\nu})\}\Bigr. 
\Bigl.\ +\ 
b_1\sum_{\substack{\mu,\nu=1\\\mu\neq\nu}}^4\{1
-\re\tr(U^{1\times2}_{x,\mu,\nu})\}\right)\,  ,
\]
with the bare inverse gauge coupling $\beta$, $b_1=-1/12$ and
$b_0=1-8b_1$. The fermionic action for two flavours of maximally
twisted, mass degenerate quarks in the so called twisted
basis~\cite{Frezzotti:2000nk,Frezzotti:2003ni} reads
\begin{equation}
  \label{eq:sf}
  S_\mathrm{tm}\ =\ a^4\sum_x\left\{ \bar\chi(x)\left[ D[U] + m_0 +
      i\mu_q\gamma_5\tau^3\right]\chi(x)\right\}\, ,
\end{equation}
where $m_0$ is the untwisted bare quark mass, $\mu_q$ is the bare
twisted quark mass, $\tau^3$ is the third Pauli matrix acting in
flavour space and
\[
D[U] = \frac{1}{2}\left[\gamma_\mu\left(\nabla_\mu +
    \nabla^*_\mu\right) -a\nabla^*_\mu\nabla_\mu \right]
\]
is the mass-less Wilson-Dirac operator. $\nabla_\mu$ and $\nabla_\mu^*$
are the forward and backward covariant difference operators,
respectively. Twisted mass fermions are said to be at \emph{maximal
 twist} if the bare untwisted mass $m_0$ is tuned to its critical
value $m_\mathrm{crit}$, the situation we shall be interested in. For 
convenience we define the hopping parameter $\kappa=1/(8+2am_0)$. Note
that we shall use the twisted basis throughout this paper.

Maximally twisted mass fermions share most of their properties with
Wilson's originally proposed formulation, but provide important
advantages: the spectrum of
$\gamma_5(D[U]+m_0+i\mu_q\gamma_5)\cdot(\textrm{h.c.})$ is bounded
from below, which was the original reason to consider twisted mass
fermions. The twisted mass $\mu_q$ is related directly to the physical
quark mass and renormalises multiplicatively only. Many mixings under
renormalisation are greatly simplified. And -- most
importantly -- as was first shown in Ref.~\cite{Frezzotti:2003ni}
physical observables are automatically $\mathcal{O}(a)$ improved
without the need to determine any operator-specific improvement
coefficients.

The main drawback of maximally twisted mass fermions is that
flavour symmetry is broken explicitly at finite value of the lattice
spacing, which amounts to $\mathcal{O}(a^2)$ effects in physical
observables. However, it turns out that 
this is presumably only relevant for the mass of the neutral pseudo
scalar meson (and closely related quantities). We shall discuss this issue
in more detail later on.

\subsection{$\mathcal{O}(a)$ Improvement in Practice}
\begin{figure}[t]
  \centering
  \subfigure[\label{fig:qr0fps}]%
  {\includegraphics[width=0.45\linewidth]{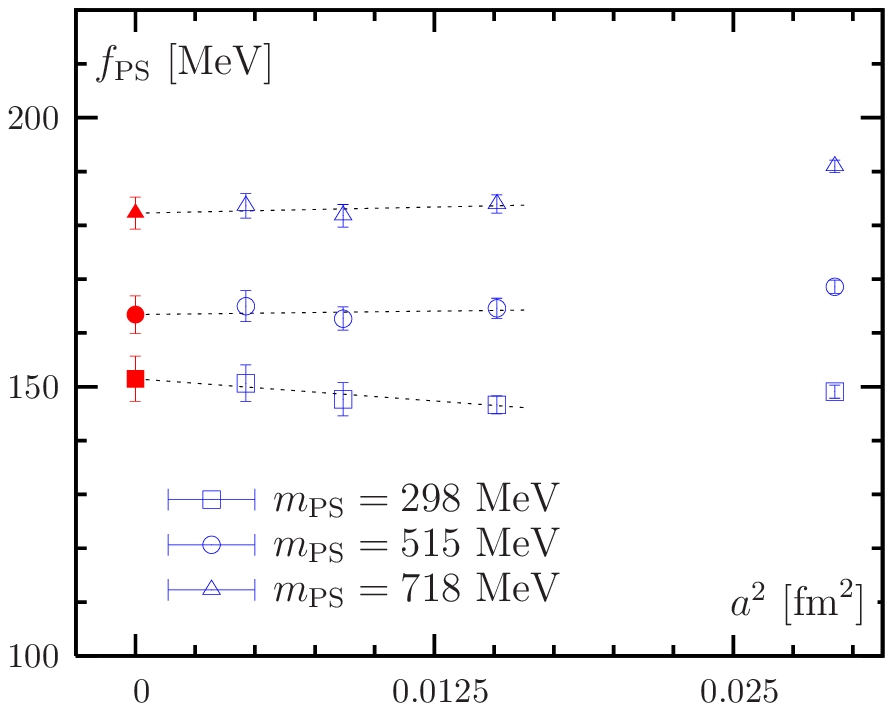}}
  \quad
  \subfigure[\label{fig:plaq}]%
  {\includegraphics[width=0.45\linewidth]{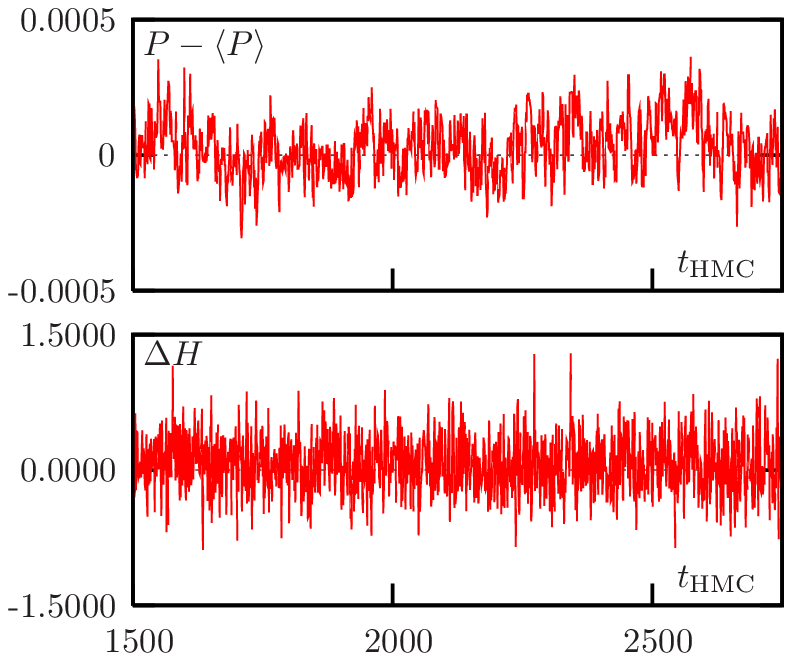}}
  \caption{(a) $f_\mathrm{PS}\ [\mathrm{MeV}]$ as a function of
    $a^2[\mathrm{fm}^2]$ in the quenched approximation \cite{Jansen:2005kk}.
  (b) Monte Carlo history of $P-\langle P\rangle$ and $\Delta H$ for
  ensemble $C_1$.}
  \label{fig:qr0fpsand}
\end{figure}

It is well established in the literature that $\mathcal{O}(a)$
improvement works very well in practice in the quenched
approximation~\cite{Jansen:2003ir,Jansen:2005gf,Jansen:2005kk,AbdelRehim:2004sp,AbdelRehim:2004gx,Abdel-Rehim:2005gz}.
As an example we show in figure~\ref{fig:qr0fps} the
\emph{essentially flat} continuum extrapolation 
in $a^2$ of the pseudo scalar decay constant $f_\mathrm{PS}$ in
physical units ($r_0=0.5\ \mathrm{fm}$ was used to set the scale) for
three reference values of the pseudo scalar mass as obtained in
Ref.~\cite{Jansen:2005kk}.

So far we did not discuss how maximal twist can be achieved in
practise and there are various solutions to this problem. Emerging from
the proof of $\mathcal{O}(a)$ improvement at maximal
twist~\cite{Frezzotti:2005gi,Shindler:2005vj} the general prescription
is to choose a parity odd 
operator $O$ and determine $am_\mathrm{crit}$ such that $O$ has
vanishing expectation value at fixed physical situation for all
lattice spacings. The physical situation can be fixed for instance by
keeping $m_\mathrm{PS}$ in physical units fixed, where
$m_\mathrm{PS}$ is the mass of the lightest charged pseudo scalar
particle. One possible quantity to tune is the PCAC quark
mass, defined as
\begin{equation}
  \label{eq:mpcac}
  m_\mathrm{PCAC} =
  \frac{\sum_\mathbf{x}\langle\partial_0A^a_0(\mathbf{x},t)P^a(0)\rangle}%
  {2\sum_\mathbf{x}\langle P^a(x)P^a(0)\rangle} \, ,\qquad\quad a=1,2\, ,
\end{equation}
where $A_\mu^a$ and $P^a$ are the axial vector current and the pseudo
scalar density, respectively,
\[
A_\mu^a(x) = \bar\chi(x)\gamma_\mu\gamma_5\frac{\tau^a}{2}\chi(x)\,
,\qquad\qquad P^a(x) = \bar\chi(x)\gamma_5\frac{\tau^a}{2}\chi(x)\, .
\]
Tuning the PCAC mass to zero has been successful in the
context of the aforementioned quenched investigations, in agreement
with theoretical considerations \cite{Sharpe:2004ny, Aoki:2004ta,
  Frezzotti:2005gi}. The collaboration follows the
strategy to determine the value of $am_\mathrm{crit}$ at the lowest
available value of $a\mu \ll a\Lambda_\mathrm{QCD}$, which is
sufficient to guarantee $\mathcal{O}(a)$ improvement
\cite{Frezzotti:2005gi}.

\section{Setup and Tuning}

In table~\ref{tab:setup} we summarise the various ensembles
produced by the ETM collaboration. We have simulations for three
different values of the inverse gauge coupling
$\beta=3.8$, $\beta=3.9$ and $\beta=4.05$. The corresponding values of
the lattice spacing are about $a\approx0.10\ \mathrm{fm}$,
$a\approx0.09\ \mathrm{fm}$ and $a\approx0.07\ \mathrm{fm}$,
respectively (see later). For each value of $\beta$ we have four or
five different values of the bare twisted mass parameter $a\mu_q$,
chosen such that the simulations cover a range of 
pseudo scalar masses between $300$ and $700\ \mathrm{MeV}$. 

The physical box sizes of the simulations at $\beta=3.9$ and
$\beta=4.05$ are roughly equal and around $2.2\ \mathrm{fm}$, while 
the volume at $\beta=3.8$ is slightly larger. For all $\beta$-values 
we have carried out simulations at different physical volumes
in order to check for finite size (FS) effects. For 
each value of $\beta$ and $\mu_q$ (ensemble) we have produced 
around $5000$ equilibrated trajectories in units of
$\tau=0.5$. The actual trajectory length $\tau$ is given in
table~\ref{tab:setup}. In all cases we allowed for at least $1500$
trajectories for equilibration (again in units of $\tau=0.5$). 

Note that the analyses for the ensembles at $\beta=3.8$ are in a very
preliminary status for reasons that will be explained later. For the
purpose of this proceeding contribution $\beta=3.8$ results are
excluded from most of the present analyses.

\begin{table}[t!]
  \centering
  \begin{tabular*}{1.0\textwidth}{@{\extracolsep{\fill}}lccccccc}
    \hline\hline
    Ensemble & $L^3\times T$ & $\beta$ & $a\mu_q$ & $\kappa$ &
    $\tau_\mathrm{int}(P)$ & $\tau_\mathrm{int}(am_\mathrm{PS})$ & $\tau$\\
    \hline\hline
    $A_1$ & $24^3\times 48$ & $3.8$ & $0.0060$ & $0.164111$ & $190(44)$
    & $8(2)$ & $1.0$\\
    $A_2$ &  & & $0.0080$ & & $172(80)$ & $10(2)$ & $1.0$\\
    $A_3$ &  & & $0.0110$ & & $130(50)$ & $6(1)$ & $1.0$\\
    $A_4$ &  & & $0.0165$ & & $40(12)$ & $6(1)$ & $1.0$\\
    $A_5$ & $20^3\times 48$ & $3.8$ & $0.0060$ & $0.164111$ & $250(100)$
    & $5(1)$ & $1.0$\\
    \hline
    $B_1$ & $24^3\times 48$ & $3.9$ & $0.0040$ & $0.160856$ & $47(15)$
    & $7(1)$ & $0.5$\\
    $B_2$ &  & & $0.0064$ &  & $23(7)$
    & $17(4)$ & $0.5$ \\
    $B_3$ &  & & $0.0085$ &  & $13(3)$ & $10(2)$ & $0.5$\\
    $B_4$ &  & & $0.0100$ &  & $15(4)$ & $7(2)$ & $0.5$\\
    $B_5$ &  & & $0.0150$ &  & $30(8)$ & $20(6)$ & $0.5$\\
    $B_6$ & $32^3\times 64$ & $3.9$ & $0.0040$ & $0.160856$ & $37(11)$
    & $2.8(3)$ & $0.5$ \\
    \hline
    $C_1$ & $32^3\times 64$ & $4.05$ & $0.003$ & $0.157010$ & $18(4)$ &
    $7(1)$ & $0.5$\\
    $C_2$ & & & $0.006$ & & $10(2)$ & $9(2)$ & $0.5$ \\
    $C_3$ & & & $0.008$ & & $13(3)$ & $7(1)$ & $0.5$\\
    $C_4$ & & & $0.012$ & & $5(1)$ & $4.8(6)$ & $0.5$\\
    $C_5$ & $24^3\times 48$ & $4.05$ & $0.006$ & $0.157010$ & $12(2)$ &
    $11(1)$ & $1.0$\\
    $C_6$ & $20^3\times 48$ & $4.05$ & $0.006$ & $0.157010$ & $10(2)$ &
    $7(1)$ & $1.0$\\
    \hline\hline
  \end{tabular*}
  \caption{Summary of ensembles produced by the ETM collaboration. We
    give the lattice volume $L^3\times T$ and the values of the
    inverse coupling $\beta$, the twisted mass parameter $a\mu_q$, the
    hopping parameter $\kappa$ and the
    trajectory length $\tau$. In addition
    we provide values for the integrated autocorrelation time of two
    typical quantities, the plaquette $P$ and the pseudo scalar mass
    $am_\mathrm{PS}$, in units of $\tau=0.5$.}
  \label{tab:setup}
\end{table}

The determination of for instance quark masses requires a
renormalisation procedure. To this end we have implemented the
non-perturbative RI-MOM renormalisation scheme
\cite{Martinelli:1994ty}, which provides in the case of maximally
twisted mass fermions $\mathcal{O}(a)$ improved renormalisation
constants \cite{etmc:rimom}. Where appropriate, we convert our results
to the $\overline{\mathrm{MS}}$ scheme at the desired scale $\mu_R$
using renormalisation group improved continuum perturbation theory at
N$^3$LO \cite{Chetyrkin:1999pq}. For details on this procedure we
refer to Ref.~\cite{etmc:rimom,Dimopoulos:2007aa}.

\begin{figure}[t]
  \centering
  \subfigure[\label{fig:wall}]%
  {\includegraphics[width=0.45\linewidth]{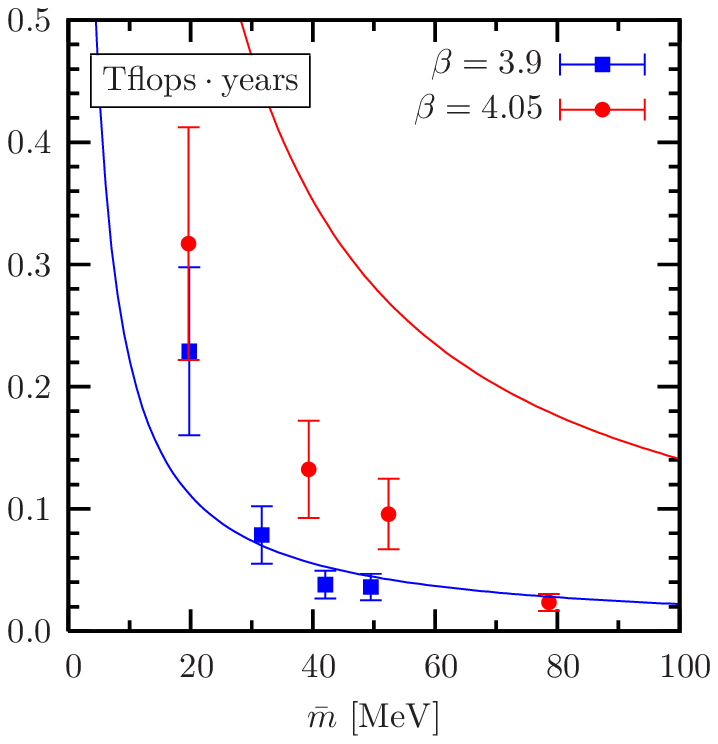}}
  \quad
  \subfigure[\label{fig:r0}]%
  {\includegraphics[width=0.43\linewidth]{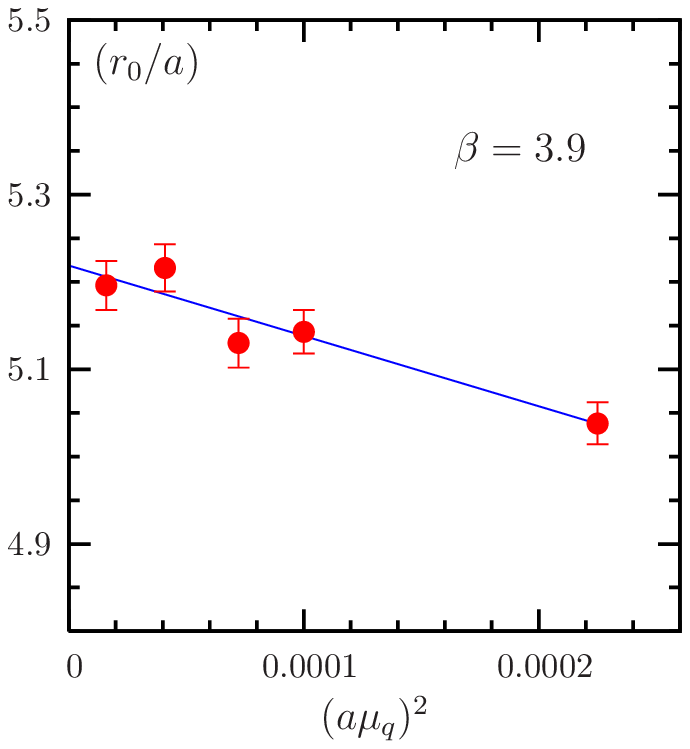}}
  \caption{(a) Comparison of the cost estimate we measure for the
    mtmHMC (data points) to the DD-HMC (lines). For the DD-HMC we used
    Eq.(3.1). The upper (lower) line compares to the $\beta=4.05$
    ($\beta=3.9$) data points.
    (b) $r_0/a$ as a function of $(a\mu)^2$ for
    $\beta=3.9$. The line represents a linear extrapolation in
    $(a\mu)^2$ to the chiral limit.
  }
  \label{fig:effmplaq}
\end{figure}

\subsection{Algorithm Stability and Performance}

The algorithm we used is a HMC algorithm \cite{Duane:1987de} with
mass preconditioning \cite{Hasenbusch:2001ne,Hasenbusch:2002ai} and
multiple time scale integration (mtmHMC), as described in detail in
Refs.~\cite{Urbach:2005ji,Jansen:2005yp}. The algorithm performs
smoothly and without any instabilities in the whole range of
simulation parameters we have available. In particular we do not
observe any instabilities at our smallest values for the twisted mass
parameter $a\mu_q$ for the different values of $\beta$. In
figure~\ref{fig:plaq} we illustrate this with the 
Monte Carlo history of the plaquette $P-\langle P\rangle$ and the
difference in the HMC Hamiltonian $\Delta H$ for ensemble $C_1$. For
further details about the algorithmic parameters used see
Ref.~\cite{etmc:long}. 

Due to the fact that the twisted mass parameter provides an infra-red
cut-off for the eigenvalue spectrum of the lattice Dirac operator we
do not expect instabilities due to very small or zero eigenvalues (for
a discussion of this issue for Wilson fermions see
Ref.~\cite{DelDebbio:2005qa}). However, Wilson type fermions exhibit a
non-trivial phase structure at finite value of the lattice spacing
with a first order phase transition at the chiral point
for our choice of the gauge and fermionic action
\cite{Farchioni:2004us,Farchioni:2004ma,Farchioni:2004fs,Farchioni:2005ec,Farchioni:2005bh,Farchioni:2005tu}.
As a consequence simulations at given, finite value of the lattice spacing
cannot be performed with arbitrarily small values of the bare quark
mass and one needs to make sure that the values of the twisted mass
parameter are large enough for the simulations not to be affected by
the first order phase transition. 

This issue was thoroughly investigated by the ETM collaboration with
the result that we do not see metastabilities for any of our
simulation points. But, as a matter of fact, simulations at
large value of the lattice spacing and maximal twist are performed
\emph{potentially} in the close vicinity of the (second order)
critical endpoint at $a\mu_c$ of the first order phase transition
line. This line extends in the $\mu_q$-$\kappa$-parameter plane to a
first approximation perpendicular to the $\kappa$-axis from
$(\kappa_\mathrm{crit},-\mu_c)$ to $(\kappa_\mathrm{crit},\mu_c)$ (see
Ref.~\cite{Shindler:2007vp} and references therein for a more detailed
discussion). Hence, there is the danger for long autocorrelation times
in quantities that are not continuous at the phase transition like for
instance the plaquette or the PCAC quark mass $m_\mathrm{PCAC}$.

This is supported by table~\ref{tab:setup} where we give estimates of
the integrated autocorrelation 
times for the plaquette and the pseudo scalar mass. While for the
pseudo scalar mass $\tau_\mathrm{int}$ is always moderately small and
depending only weakly on the values of $\beta$ and $\mu_q$,
this is not the case for the plaquette: there is a trend in the data
that $\tau_\mathrm{int}(P)$ decreases with increasing values of $\mu_q$
and $\beta$, even though the statistical errors are so large
that the $\mu_q$-dependence is not significant. Note that the
$\tau_\mathrm{int}$-values for 
$m_\mathrm{PCAC}$ are similar to those for the plaquette.
Fortunately, we observe very long autocorrelation times only for our
smallest $\beta$-value. This is the reason that we are still
investigating the error analysis for those ensembles. 
This affects the determination of the PCAC quark mass, which is 
needed to tune to maximal twist. Hence, at this stage, we use the
corresponding results at $\beta=3.8$ only for estimating systematic
errors.

In order to investigate algorithm stability it was suggested in
Ref.~\cite{Giusti:2007hk} to study the statistical 
distribution of the twist angle as a function of the quark mass. The
quantity of interest is the PCAC quark mass, evaluated in the
background of a given gauge configuration by taking the axial to pseudo
scalar correlator at some large value for $t/a$ normalised by
$am_\mathrm{PS}/(2C_\mathrm{PP})$. Here $C_\mathrm{PP}$ is the pseudo
scalar correlator averaged over all gauges evaluated at the same value
of $t/a$. The PCAC quark mass is related to the twist angle $\omega$
by 
\[
\tan\omega = \frac{\mu_q}{Z_\mathrm{A}m_\mathrm{PCAC}}\, .
\]
The statistical distribution of $am_\mathrm{PCAC}$ resembles always to
a very good approximation a normal distribution. As will be discussed
later, at fixed value of $\beta$and $L/a$ the distribution mean
depends on $a\mu_q$ as expected. For all three $\beta$-values the standard 
deviation is increasing at the order of a few percent with increasing
$a\mu_q$. In addition we observe a smooth dependence of the
standard deviation on $L/a$ and the lattice spacing. Hence, there is
no sign for instabilities seen in $\tan\omega$. 

It is interesting to compare the performance of our HMC variant with
the one using domain decomposition as a preconditioner (DD-HMC)
(instead of mass preconditioning), which is described in
Ref.~\cite{Luscher:2005rx}. A useful performance figure is the number
of floating point operations $C_\mathrm{op}$ required to generate
$1000$ independent gauge configurations as a function of the box size
$L [\mathrm{fm}]$, the lattice spacing $a [\mathrm{fm}]$ and the quark
mass $\bar{m} [\mathrm{MeV}]$ in the $\overline{\mathrm{MS}}$ scheme
at $\mu_R=2\ \mathrm{GeV}$  
\cite{Ukawa:2002pc,DelDebbio:2006cn}
\begin{equation}
  \label{eq:cost}
  C_\mathrm{op} = k\ \left( \frac{20\ \mathrm{MeV}}{\bar{m}}\right)^{c_m}\
  \left(\frac{L}{3\ \mathrm{fm}}\right)^{c_L}\ \left(\frac{0.1\
      \mathrm{fm}}{a}\right)^{c_a}\
  \mathrm{Tflops}\times\mathrm{years}\ .
\end{equation}
In Ref.~\cite{DelDebbio:2006cn} the parameters in Eq.~(\ref{eq:cost})
for the DD-HMC algorithm with Wilson fermions are estimated roughly to
$k=0.3$, $c_m=1$, $c_L=5$ and $c_a=6$, which is a significant
improvement as compared to cost estimates for the original HMC
algorithm, see for instance Ref.~\cite{Ukawa:2002pc}. Using the 
integrated autocorrelation time of the plaquette as a measure for the
autocorrelation of two gauge configurations we can measure
$C_\mathrm{op}$ for the mtmHMC algorithm and compare the result in
figure~\ref{fig:wall} to the cost estimate for the DD-HMC
algorithm~\cite{DelDebbio:2006cn}.

Figure~\ref{fig:wall} reveals that the performance of the two
algorithms is very similar. In particular for the larger value of the
two plotted lattice spacings the agreement is rather good. For our HMC
version the lattice spacing dependence appears to be milder. However,
one should keep in mind the large errors associated to this cost
figure. Moreover, the result may depend significantly on how much
effort is invested into tuning of algorithmic parameters.

\subsection{Sommer Parameter $r_0$}

In order to be able to compare results at different values of the
lattice spacing it is convenient to measure the hadronic scale $r_0/a$
\cite{Sommer:1993ce}. It is defined via the force between static
quarks at intermediate distance and can be measured to high accuracy
in lattice QCD simulations. For details on how we measure $r_0/a$ we
refer to Ref.~\cite{etmc:long}.

In figure~\ref{fig:r0} we show $r_0/a$ as a function of $(a\mu_q)^2$ for
$\beta=3.9$. The mass dependence appears to be rather weak and a
quadratic fit in $a\mu_q$ to our data describes the data rather
well.  The results we obtain for the three $\beta$-values in the
chiral limit are
$r_0/a=4.46(3)$ at $\beta=3.8$, $r_0/a=5.22(2)$ at $\beta=3.9$ and
$r_0/a=6.61(3)$ at $\beta=4.05$. The statistical accuracy
for $r_0/a$ is about $0.5\%$. Note that the data is also compatible
with a linear dependence on $a\mu_q$ and a linear fit gives consistent
results in the chiral limit.

\begin{figure}[t]
  \centering
  \subfigure[\label{fig:mpcac}]%
  {\includegraphics[width=0.45\linewidth]{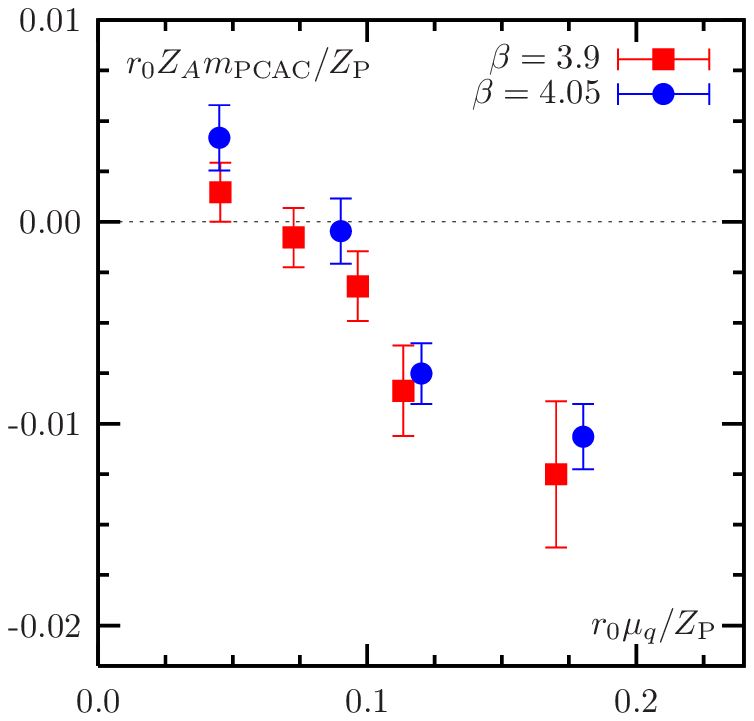}}
  \quad
  \subfigure[\label{fig:effm}]%
  {\includegraphics[width=0.45\linewidth]{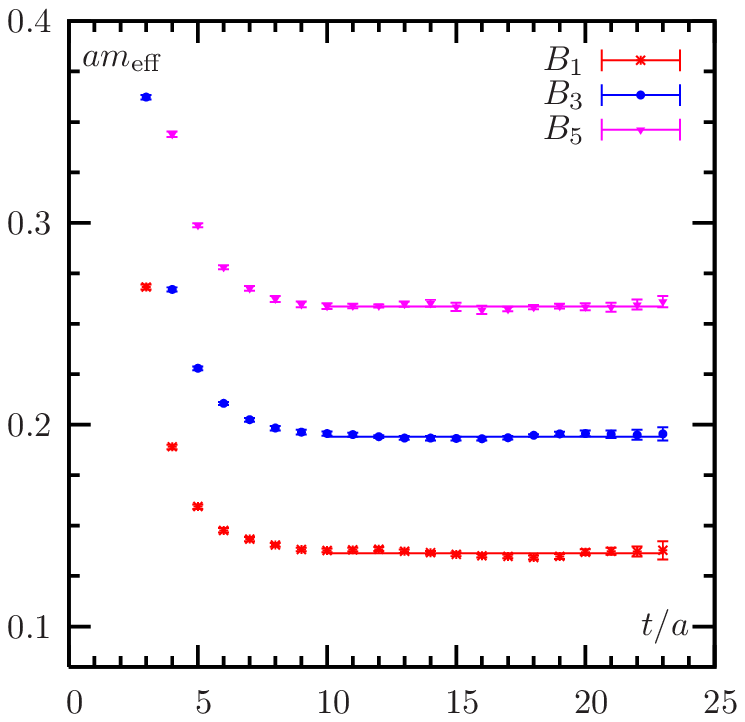}}
  \caption{(a) renormalised PCAC quark mass against renormalised twisted mass
    for $\beta=3.9$ and $\beta=4.05$. The statistical uncertainty on
    $Z_\mathrm{P}$ is not included.
    (b) Effective mass plot for the charged pseudo scalar mass for
    ensemble $B_1$, $B_3$ and $B_5$. The lines represent the fitted
    value for $am_\mathrm{PS}$.
  }
  \label{fig:c}
\end{figure}

\subsection{Tuning to Maximal Twist}

In order to obtain $\mathcal{O}(a)$ improvement the bare quark mass
$m_0$ must be tuned to its critical value. As mentioned before we use
the PCAC quark mass defined in Eq.~(\ref{eq:mpcac}) for tuning to
maximal twist. The goal was to tune this quantity to zero
approximately at the smallest available $\mu_q$-value at each lattice
spacing, which corresponds to approximately fixed physical pseudo
scalar mass. Considering only $\beta=3.9$ and $\beta=4.05$, it was
possible to perform this tuning task with two or three tuning runs for
each lattice spacing.

The result is plotted in figure~\ref{fig:mpcac}, where we plot the
renormalised $m_\mathrm{PCAC}$ against the renormalised $\mu_q$, both in
units of $r_0$, for $\beta=3.9$ and $\beta=4.05$. The renormalisation
factors were determined using the RI-MOM scheme. Within the statistical
accuracy the PCAC quark mass is zero at a common value of the
renormalised twisted mass. For all other values of $\mu_q$ we observe
(small) deviations from zero. This $\mu_q$ dependence is an
$\mathcal{O}(a)$ cut-off effect which will modify \emph{only} the
$\mathcal{O}(a^2)$ lattice artifacts of physical observables.
The numerical precision we were aiming for was
$a\Lambda_\mathrm{QCD}\epsilon/\mu\lesssim0.01$, where $\epsilon$ is
the uncertainty on the PCAC quark mass at the lowest value of
$\mu_q$~\cite{roberto}. It turns out that the accuracy we have
achieved is sufficient for excellent scaling properties, as will be
seen later.

\section{Results}

The ETM collaboration is currently analysing the generated gauge
configurations for many different observables. Not all of these --
partly preliminary -- results can be summarised here and we shall
therefore mainly concentrate on the pseudo scalar sector. At the end
of this section we present an overview of the available results and
give references of the corresponding proceedings contributions.

\subsection{Charged Pseudo Scalar Mass and Decay Constant}

The charged pseudo scalar meson mass $am_\mathrm{PS}$ can be extracted
from the time exponential decay of suitable correlation functions. For
details on our analysis procedure see Ref.~\cite{etmc:long}. To
demonstrate the quality of our data we show in figure~\ref{fig:effm}
effective mass plots for the ensembles $B_1$, $B_3$ and $B_5$ obtained
from the pseudo scalar correlation function. The final values for the
masses are obtained from a fit to a $4\times4$ matrix of correlators
\cite{etmc:long}. We also attempted to determine the energy of the
first excited state of the pseudo scalar meson. We were not able to
determine it with any reliability from an unconstrained fit. (In particular,
there is no evidence for an excited state with mass
$m_\mathrm{PS}^\pm+m_\mathrm{PS}^0$, which is theoretically possible
for maximally twisted mass fermions~\cite{Shindler:2007vp}.)
Constraining the energy of the first excited state to three times the
ground state mass, however, does allow for an acceptable fit.

\begin{figure}[t]
  \centering
  \subfigure[\label{fig:fpsmpisq}]%
  {\includegraphics[width=0.43\linewidth]{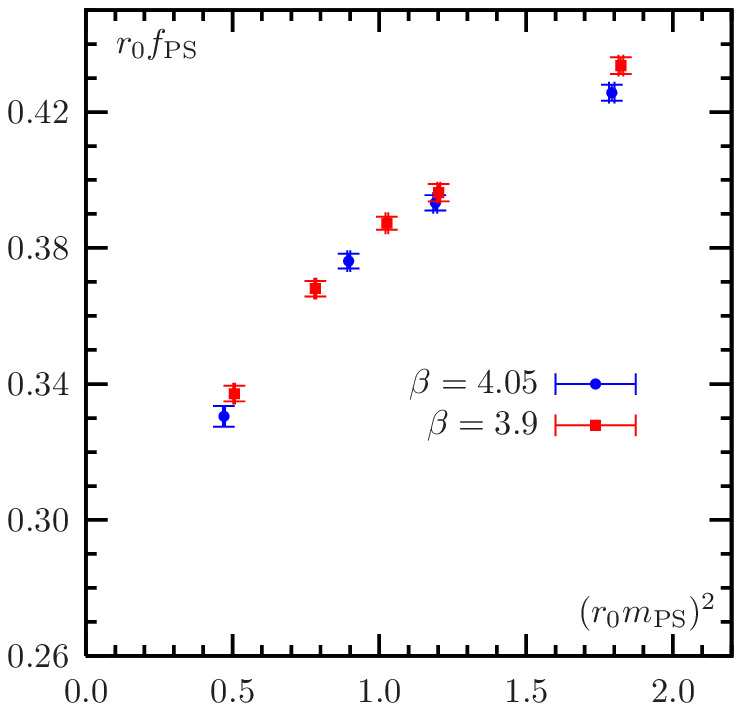}}
  \quad
  \subfigure[\label{fig:r0fps}]%
  {\includegraphics[width=0.45\linewidth]{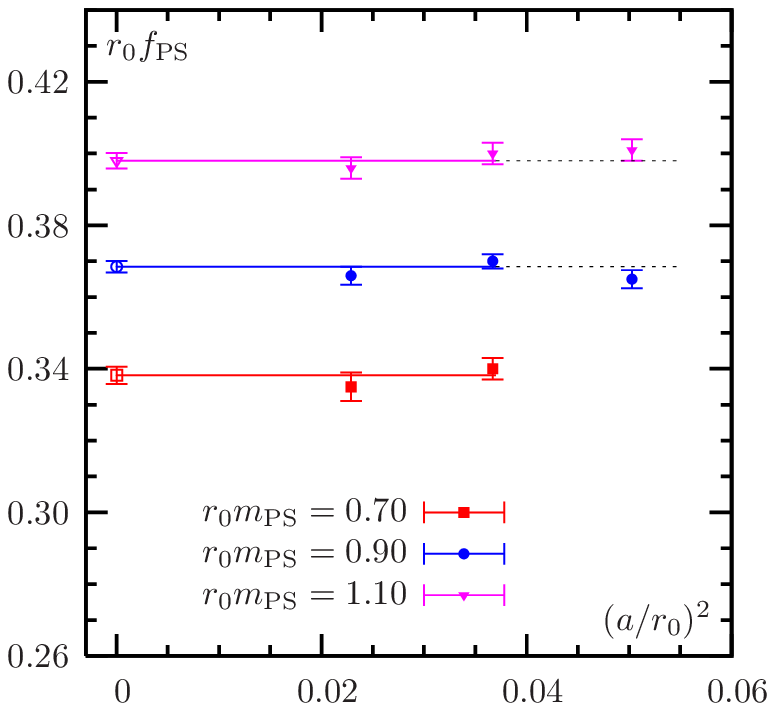}}
  \caption{(a) $r_0f_\mathrm{PS}$ as a function of
    $(r_0m_\mathrm{PS})^2$ for $\beta=3.9$ ($B_1$ to $B_5$) and
    $\beta=4.05$ ($C_1$ to $C_4$). 
    (b) Continuum extrapolation of $f_\mathrm{PS}$ at fixed volume for
    three reference values of $r_0m_\mathrm{PS}$. The data points at
    $\beta=3.8$ are not used.
  }
  \label{fig:fps}
\end{figure}

For maximally twisted mass fermions the charged pseudo scalar decay
constant $af_\mathrm{PS}$ can be extracted from 
\begin{equation}
  \label{eq:fps}
  f_\mathrm{PS} = \frac{2\mu_q}{m_\mathrm{PS}^2} | \langle
  0|P^a|\pi\rangle|\, ,\qquad a=1,2\, 
\end{equation}
due to the exact lattice PCVC relation with no need to compute any
renormalisation constant \cite{Frezzotti:2000nk}. Thanks to this
advantage and having the results for $r_0/a$ at hand we plot the
results for $r_0f_\mathrm{PS}$ as a function of $(r_0m_\mathrm{PS})^2$
for $\beta=3.9$ and $\beta=4.05$ in figure~\ref{fig:fpsmpisq}. We plot
only the results for the ensembles $B_1$ to $B_5$ and $C_1$ to $C_4$
in order to have approximately equal physical volumes for the two
values of $\beta$. Figure~\ref{fig:fpsmpisq} provides first evidence
that lattice artifacts in those two quantities are small. 

\subsubsection*{Continuum Extrapolation of $f_\mathrm{PS}$ at Fixed Volume}

This statement can be brought to a more quantitative level for
instance for the pseudo scalar decay constant. To this end we
interpolate $af_\mathrm{PS}$ at every lattice spacing linearly in
$(am_\mathrm{PS})^2$ to fixed values of $r_0m_\mathrm{PS}$. We
corrected for the very small difference in the physical
volume~\cite{roberto}.

The interpolated data points are plotted in figure~\ref{fig:r0fps}
for three values of $r_0m_\mathrm{PS}=0.70$, $r_0m_\mathrm{PS}=0.90$ and
$r_0m_\mathrm{PS}=1.10$ as functions of $(a/r_0)^2$. It is visible
that the differences between the data at $\beta=3.9$ and $\beta=4.05$
are of the order of the statistical accuracy and hence, we perform
a weighted average of these two lattice spacings to obtain continuum
estimates. This indicates again that scaling violations are very small
in the pseudo scalar decay constant.

Even though we are still investigating the error analysis we plot in
figure~\ref{fig:r0fps} also the results for $\beta=3.8$ for those
reference points where we are able to interpolate to the reference
values of $r_0m_\mathrm{PS}$ (for the lowest value of
$r_0m_\mathrm{PS}$ we currently need to extrapolate). Keeping in
mind that these data points are very preliminary we can nevertheless
say that they fit rather well into the picture that lattice artifacts
are very small.

Other quantities -- for instance the renormalised quark mass --
show a similar scaling behaviour as discussed in
Ref.~\cite{roberto},  where also the
results at $\beta=3.8$ are discussed in more detail.

\subsubsection*{Finite Size Effects}

\begin{figure}[t]
  \centering
  \subfigure[\label{fig:FS}]%
  {\includegraphics[width=0.45\linewidth]{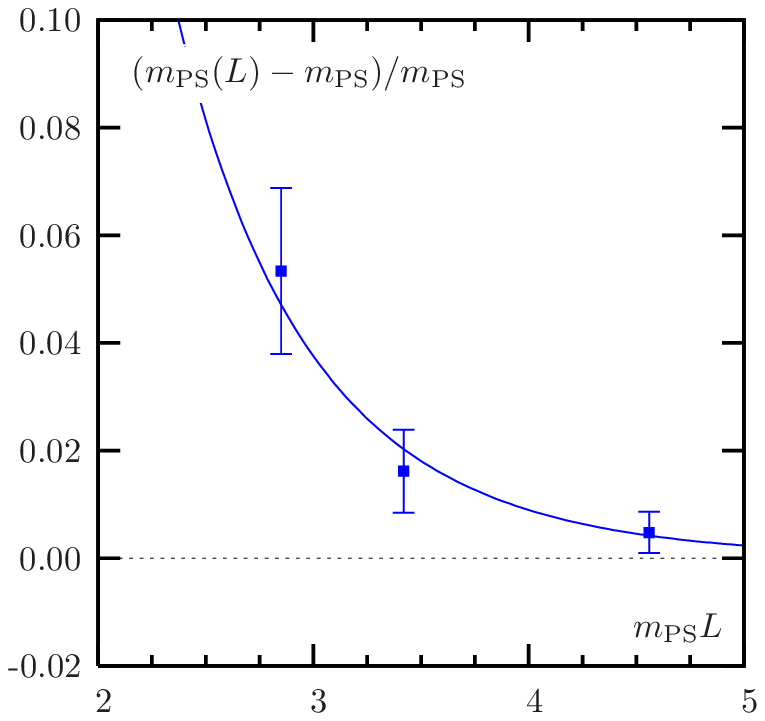}}
  \quad
  \subfigure[\label{fig:fpschi}]%
  {\includegraphics[width=0.43\linewidth]{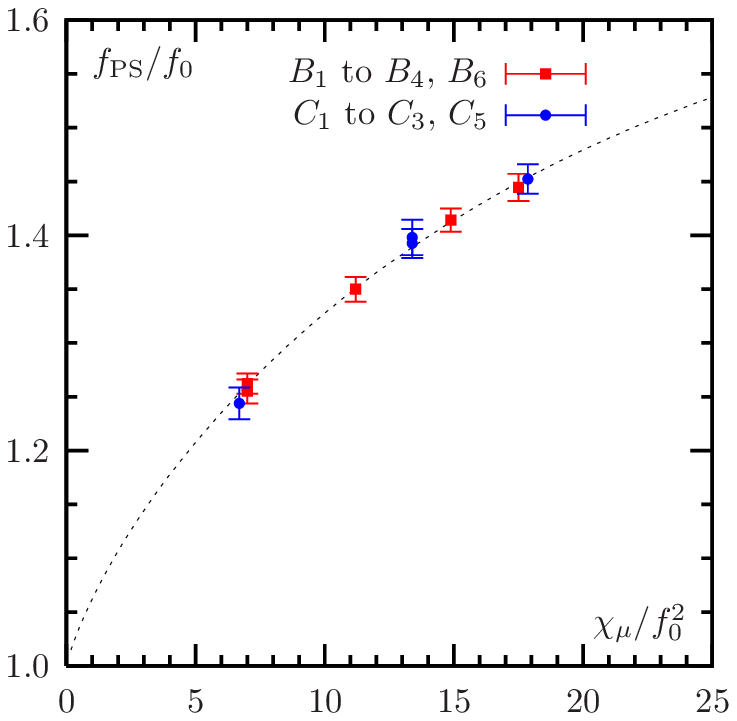}}
  \caption{(a) Relative finite size effects for ensembles $C_2$, $C_5$
    and $C_6$. The line represents a fit with formula~
    (\protect\ref{eq:fit}) to our data.
    (b) $f_\mathrm{PS}/f_0$ as a function of $\chi_\mu/f_0^2$ as obtained
    from a combined fit to $\chi$PT formulae. The dashed line is the
    fitted function Eq.~(\protect\ref{eq:chirfo2}), and
    the data points are FS corrected.
  }
  \label{fig:a}
\end{figure}

At the level of statistical accuracy we have achieved now, finite size
effects for $f_\mathrm{PS}$ and $m_\mathrm{PS}$ cannot be neglected.
It is therefore of importance to study whether the FS effects
can be described within the framework of chiral perturbation
theory. This requires to compare simulations with different lattice
volumes and all other parameters kept fixed, like for instance
ensembles $C_2$, $C_5$ and $C_6$ or $B_1$ and $B_6$. For all these
ensembles $m_\mathrm{PS}L\geq 3$ holds, which is believed to be needed
for $\chi$PT formulae to apply. The first observation we make is that
the finite size effects are compatible with an exponential behaviour
in $m_\mathrm{PS}L$. As an example we plot in figure~\ref{fig:FS} the
relative finite size effects 
\begin{equation}
  \label{eq:R}
  R_O \equiv \frac{O(L) - O(L=\infty)}{O(L=\infty)}
\end{equation}
for $O=m_\mathrm{PS}$ against $m_\mathrm{PS}L$. The value of
$m_\mathrm{PS}\equiv m_\mathrm{PS}(L=\infty)$ was obtained by fitting
the chiral perturbation theory inspired formula
\begin{equation}
  \label{eq:fit}
  m_\mathrm{PS}(L) = m_\mathrm{PS} + \alpha e^{-m_\mathrm{PS}L}/L^{3/2}
\end{equation}
to our data, with $\alpha$ and $m_\mathrm{PS}$ being free parameters
\cite{Guagnelli:2004ww}. A similar fit can be performed for
$f_\mathrm{PS}$ by using the value of $m_\mathrm{PS}$ as obtained from
the first fit as input. Both fits describe the data rather well with 
$\chi^2/\mathrm{dof} \ll 1$.

\begin{table}[t!]
  \centering
  \begin{tabular*}{.8\textwidth}{@{\extracolsep{\fill}}lccccc}
    \hline\hline
    & Ensemble & $m_\mathrm{PS}L$ & $R^\mathrm{meas}$ & $R^\mathrm{GL}$
    & $R^\mathrm{CDH}$ \\ 
    \hline\hline
    $m_\mathrm{PS}$ & $B_1$ & $3.3$ & $+1.8$ & $+0.62$ &
    $+1.0$ \\
    $f_\mathrm{PS}$ & $B_1$ & $3.3$ & $-2.5$ & $-2.5$ &
    $-2.4$ \\
    \hline
    $m_\mathrm{PS}$ & $C_5$ & $3.5$ & $+1.1$ & $+0.8$ & $+1.3$\\
    $f_\mathrm{PS}$ & $C_5$ & $3.5$ & $-1.8$ & $-3.2$ & $-2.9$\\
    \hline
    $m_\mathrm{PS}$ & $C_6$ & $3.0$ & $+6.2$ & $+1.8$ & $+4.7$\\
    $f_\mathrm{PS}$ & $C_6$ & $3.0$ & $-10.7$ & $-7.3$ & $-8.9$\\
    \hline\hline
  \end{tabular*}
  \caption{Comparison of measured relative FS effects
    Eq.~(\protect\ref{eq:R}) in $\%$ to
    estimates from $\chi$PT formulae.}
  \label{tab:R}
\end{table}

We shall now compare the \emph{measured} finite size effects
to predictions of \emph{continuum} chiral perturbation theory
($\chi$PT) (lattice artifacts appear to be negligible, as discussed
before). The comparison reveals that continuum $\chi$PT formulae can
describe FS effects within our statistical accuracy.

The NLO $\chi$PT formulae for $m_\mathrm{PS}$ and
$f_\mathrm{PS}$ were derived in Ref.~\cite{Gasser:1986vb} (for short
GL) and can be written as  
\begin{equation}
  \label{eq:fs}
  \begin{split}
    m_\mathrm{PS}(L) &=
    m_\mathrm{PS}\Bigl[1+\frac{1}{2}\xi\tilde{g}_1(\lambda)\Bigr]\equiv
    m_\mathrm{PS}\ K_m^{\mathrm{GL}}(L)\, ,\\
    f_\mathrm{PS}(L) &= 
    f_\mathrm{PS}\ \Bigl[1-2\xi\tilde{g}_1(\lambda)\Bigr]\ \equiv
    f_\mathrm{PS}\ K_f^\mathrm{GL}(L)\, ,
  \end{split}
\end{equation}
where
\begin{equation}
  \label{eq:xi}
  \xi = m_\mathrm{PS}^2/(4\pi f_0)^2\, ,\qquad\lambda = m_\mathrm{PS} L\, ,
\end{equation}
$\tilde{g}_1$ is a known function \cite{Gasser:1986vb} and the finite
size corrections $K_{m,f}^\mathrm{GL}$ depend apart from $L$ and
$m_\mathrm{PS}$ only on the unknown leading order low energy constant
$f_0$ (note that our normalisation is such that $f_\pi=130.7\
\mathrm{MeV}$). For the pseudo scalar meson mass the corrections are
also known to two 
loops~\cite{Colangelo:2006mp}, but the asymptotic L{\"u}scher
formula~\cite{Colangelo:2003hf,Colangelo:2005gd} (for short CDH)
provides an easier way to access higher order corrections to
$m_\mathrm{PS}$ and $f_\mathrm{PS}$, and the differences to the NNLO
formula turn out to be small~\cite{Colangelo:2006mp}. The drawback of
CDH compared to GL is that additional parameters are needed as an
input, among others the low energy constants $\Lambda_1$, $\Lambda_2$,
$\Lambda_3$ and $\Lambda_4$. Since we do not have enough data points to
determine all these parameters from a fit to only FS data we have to
rely on estimates available in the literature~\cite{Colangelo:2005gd}.

Assuming that the results for the ensembles $B_6$ and $C_2$ provide
the infinite volume estimates for $m_\mathrm{PS}$ and $f_\mathrm{PS}$,
we can compute the relative FS effects 
using Eq.~(\ref{eq:R}) for the ensembles $B_1$, $C_5$ and $C_6$, which
we denote with $R^\mathrm{meas}$. These measured estimates can then be
compared to the $\chi$PT predictions computed with the formulae from GL
and CDH, denoted by $R^\mathrm{GL}$ and $R^\mathrm{CDH}$, respectively.
The values of the unknown low energy constants are set to the estimates
provided in Ref.~\cite{Colangelo:2005gd}. In order to do so and for
evaluating the CDH formulae we need the value of the lattice spacing
which we estimate using $r_0=0.45\ \mathrm{fm}$.

The results for $R^\mathrm{meas}$, $R^\mathrm{GL}$ and
$R^\mathrm{CDH}$ are compiled in table~\ref{tab:R}. It turns out that
-- in particular for $m_\mathrm{PS}$ -- the asymptotic formula from
CDH describes the data better than the one loop formula from
GL: the CDH corrected data is in agreement with the infinite
volume estimate within the statistical accuracy. Note that the
observation  that GL usually strongly underestimates the FS effects in
$m_\mathrm{PS}$ can also be made from the Wilson data published in
Ref.~\cite{Orth:2005kq} (cf. Ref.~\cite{Giusti:2007hk}).

These results make us confident that our simulations have eventually
reached a regime of pseudo scalar masses and lattice volumes where
$\chi$PT formulae can be used to estimate FS effects. But
it is clear that in particular the CDH formula is affected by
large uncertainties, mainly stemming from the only poorly known low
energy constants, which are needed as input. Changing their values in
the range suggested in Ref.~\cite{Colangelo:2005gd}, however, changes
the estimated finite size effects maximally at the order of about
$20\%$ (of the corrections themselves).

\subsubsection*{Quark Mass Dependence and Chiral Perturbation Theory}
\label{subsec:fit}

\begin{figure}[t]
  \centering
  \subfigure[\label{fig:mpschi}]%
  {\includegraphics[width=0.43\linewidth]{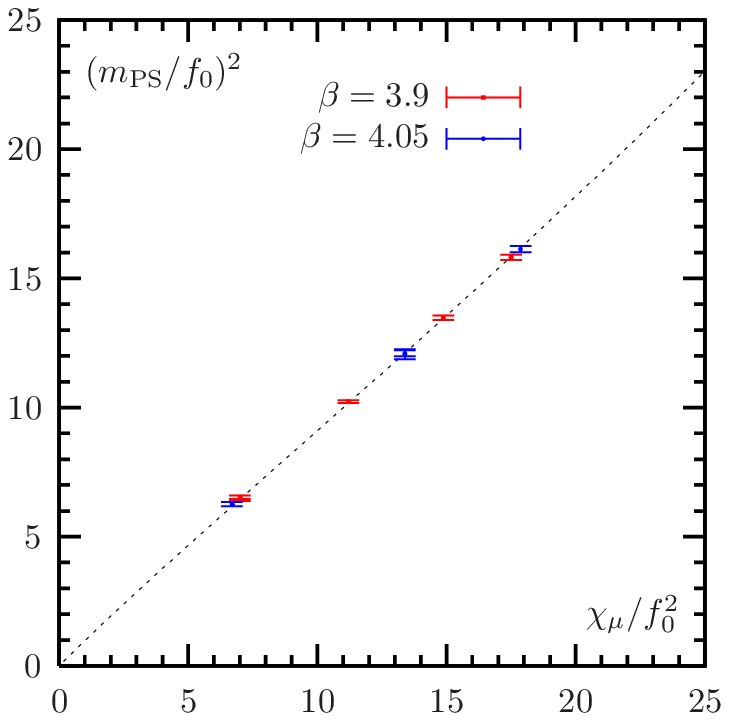}}
  \quad
  \subfigure[\label{fig:mpsmuchi}]%
  {\includegraphics[width=0.45\linewidth]{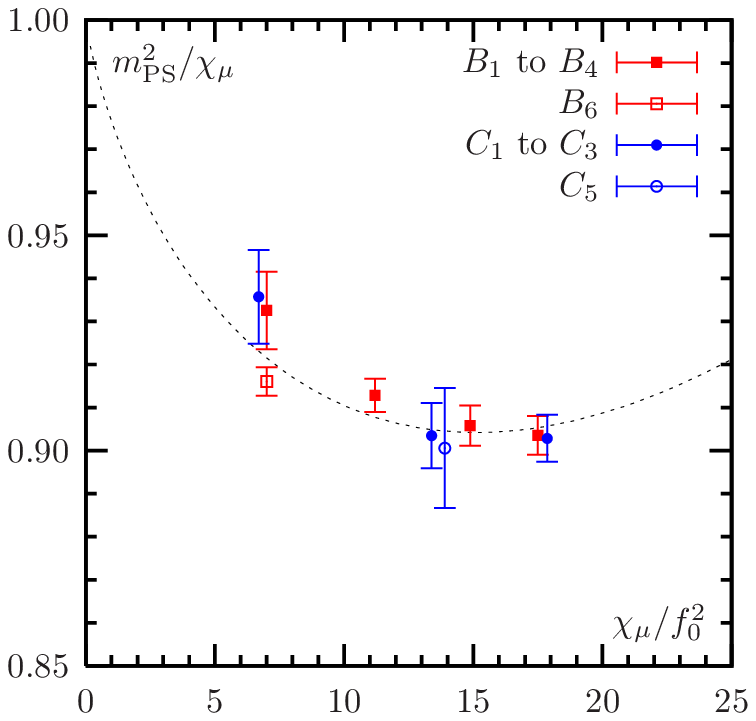}}
  \caption{(a) $(m_\mathrm{PS}/f_0)^2$ as a function of $\chi_\mu/f_0^2$
    (b) $m_\mathrm{PS}^2/\chi_\mu$ as a function of $\chi/f_0^2$. The
    data point for ensemble $C_5$ is slightly displaced. In (a) and
    (b) the dashed line represents the fitted functions
    Eqs.~(\protect\ref{eq:chirfo1}) and (\protect\ref{eq:chirfo2}),
    respectively, and the data points are FS corrected. 
  }
  \label{fig:b}
\end{figure}

So far we have argued that lattice artifacts in charged $m_\mathrm{PS}$ and
$f_\mathrm{PS}$ for $\beta=3.9$ and $\beta=4.05$ are very small and
that FS effects can be described using $\chi$PT formulae. We shall now
show that also the quark mass dependence of these quantities can
be successfully described using $\chi$PT. This will in addition allow
the determination of most of the aforementioned low energy constants
and the lattice scale.

As chiral symmetry is broken by the lattice Wilson term, eventually
the chiral extrapolation should be done using continuum extrapolated,
infinite volume data. This is described for our data in
Ref.~\cite{roberto}. However, since lattice artifacts are not visible 
with our current statistical accuracy of about $1\%$ (and chiral
symmetry breaking as well as flavour symmetry breaking are formally
lattice artifacts of $\mathcal{O}(a^2)$), we can follow a different
approach.

This approach consists of describing our data for $f_\mathrm{PS}$ 
and $m_\mathrm{PS}$ at $\beta=3.9$ and $\beta=4.05$ simultaneously
with \emph{continuum} chiral perturbation theory. We fit the
appropriate ($N_f=2$) continuum NLO $\chi$PT
formulae~\cite{Gasser:1986vb,Colangelo:2005gd} 
\begin{equation}
  m_\mathrm{PS}^2(L,\mu_q) = \chi_\mu \ K_m^2(L)\ \left[ 1 +
    \xi \log ( \chi_\mu/\Lambda_3^2 ) \right] \, ,
  \label{eq:chirfo1}
\end{equation}
\begin{equation}
  f_\mathrm{PS}(L,\mu_q)\ \ = f_0 \ \ K_f(L)\ \left[ 1 -
    2 \xi \log ( \chi_\mu/\Lambda_4^2 ) \right] \, ,
  \label{eq:chirfo2}
\end{equation}
to our raw data for $m_\mathrm{PS}$ and $f_\mathrm{PS}$ simultaneously
for $\beta=3.9$ and $\beta=4.05$. $K_{m,f}(L)$ parametrise the FS
corrections for which we can either use the GL or the CDH formulae. Both
$K_m$ and $K_f$ depend on low energy constants as well as on $L$ and
$m_\mathrm{PS}$. The notation is 
\begin{equation}
  \chi_\mu=2\hat{B}_0Z_\mu\mu_q\, ,\qquad\xi = \chi_\mu/(4\pi f_0)^2 \, ,
\end{equation}
and the normalisation $f_0=\sqrt{2}F_0$, i.e. $f_\pi=130.7\
\mathrm{MeV}$. In Eqs.~(\ref{eq:chirfo1}) and~(\ref{eq:chirfo2}) 
NNLO $\chi$PT corrections are assumed to be negligible. This approach
has the advantage that we can include finite size data consistently in
the fit and that we can use more raw data points. In addition we do
not need to interpolate our data to reference points. The fit
presented in the following is an extension to the fit presented for
only $\beta=3.9$ in Ref.~\cite{Boucaud:2007uk}.

The fit can be parametrised by six free parameters: two dimensionless
ratios $\Lambda_{3,4}/f_0$ and $f_0$ and $Z_\mu\hat{B}_0$ in
lattice units for both $\beta$-values,
i.e. $\Lambda_3/f_0$, $\Lambda_4/f_0$, $af_0|_{\beta=3.9}$,
$af_0|_{\beta=4.05}$, $aZ_\mu\hat{B}_0|_{\beta=3.9}$ and
$aZ_\mu\hat{B}_0|_{\beta=4.05}$.

Finite size effects are corrected for by using the asymptotic formulae
from CDH, which is consistently included in the fit. The parameters
that are not fitted, basically $\Lambda_1$ and $\Lambda_2$, are set to
the values suggested in Ref.~\cite{Colangelo:2005gd}. Also the lattice
spacing in $\mathrm{fm}$, needed for evaluating the CDH formula,  can
be determined consistently from the fit by setting
$f_\mathrm{PS}=f_\pi$ where the ratio $m_\mathrm{PS}/f_\mathrm{PS}$
assumes its physical value. Note that in this ratio we use the
physical value of the neutral pseudo scalar meson mass, in order to
account for electro magnetic effects not present in the lattice
simulation.

The ensembles we use in the fit are $B_1$ to $B_4$ and $C_1$ to
$C_3$. In addition we include the ensembles $B_6$ and $C_5$ in the fit
in order to explore the $L$ dependence of our data. We do not use
$C_6$ in order not to give too much weight to this $\mu_q$-value. The
ensembles $B_5$ and $C_4$, the largest available masses, are not
included in the fit, because they lead to significantly increased
$\chi^2/\textrm{dof}$ in the fit and hence we conclude that NLO 
$\chi$PT is not appropriate for such large mass-values.

Our data for $f_\mathrm{PS}$ and $m_\mathrm{PS}$ is described excellently
by the $\chi$PT formulae, see figures~\ref{fig:fpschi},
\ref{fig:mpschi} and \ref{fig:mpsmuchi}, where we plot appropriate,
dimensionless ratios. The fitted values of the parameters are
\begin{equation}
  \label{eq:fitpars}
  \begin{split}
  \Lambda_3/f_0 = 6.41(26)\, ,\quad & \quad \Lambda_4/f_0 = 11.51(21)\, ,\\
  af_0|_{\beta=3.9} = 0.0527(4)\, ,\quad & \quad af_0|_{\beta=4.05} = 0.0411(4)\, ,\\
  2aZ_\mu\hat{B}_0|_{\beta=3.9} = 4.87(4)\, ,\quad &\quad
  2aZ_\mu\hat{B}_0|_{\beta=4.05} = 3.76(3)\, ,\\
  \end{split}
\end{equation}
with $\chi^2/\mathrm{dof}=12/12$ and statistical errors 
only. The sensitivity to $\Lambda_{3,4}$ is visualised by the
deviation from linearity in figures~\ref{fig:fpschi} and
\ref{fig:mpsmuchi}.

The statistical uncertainties are estimated using a bootstrap
procedure: at each of our data points we produced $1000$ bootstrap
samples of $f_\mathrm{PS}$ and $m_\mathrm{PS}$ and used them to
perform $1000$ fits. The statistical uncertainty is then given by the
variance over these $1000$ fit results. Note that this procedure
automatically takes into account the cross-correlation between
$f_\mathrm{PS}$ and $m_\mathrm{PS}$ for a given ensemble. 

Systematic uncertainties are -- as usual -- harder to estimate. We
include the effects coming from (a) finite size effects by using GL
instead of CDH to estimate finite volume effects, (b) finite size
effects by including $\Lambda_1/f_0$ and  $\Lambda_2/f_0$ as free
parameters in the fit and (c) lattice artifacts by performing the fits
separately for $\beta=3.9$ and $\beta=4.05$.
They are discussed in more detail in Ref.~\cite{roberto}.

\subsubsection*{Lattice Calibration and Low Energy Constants}

As mentioned above, by fixing the value of $f_\mathrm{PS}$ to the
physical value of the pion decay constant $f_\pi=130.7\ \mathrm{MeV}$
where $m_\mathrm{PS}/f_\mathrm{PS}$ assumes its physical value we can
calibrate our lattices for the two $\beta$-values, with the result
\begin{equation}
  \label{eq:as}
  a|_{\beta=3.9} = 0.0855(5)\ \mathrm{fm}\, ,\qquad a|_{\beta=4.05} =
  0.0667(5)\ \mathrm{fm}\, .
\end{equation}
The ratio $a|_{\beta=3.9}/a|_{\beta=4.05}=1.28(1)$ can be compared with the ratio
$(r_0/a)|_{\beta=4.05}/(r_0/a)|_{\beta=3.9}=1.27(1)$ and we find excellent
agreement. Moreover, using those values for the lattice spacings we
can get an estimate of the ratio $Z_\mu|_{\beta=3.9}/Z_\mu|_{\beta=4.05}$
from our fit and compare it to the corresponding ratio as determined
with RI-MOM. Also here the agreement is excellent. We take this as an
indication that the combined fit does not hide lattice artifacts in
fit parameters.

Having set the scale allows us to determine the low energy constants
\begin{equation}
  \label{eq:lbars}
  \bar\ell_3 \equiv\log\left(\frac{\Lambda_3^2}{m_{\pi}^2}\right) =
  3.44(8)(35) 
  \, ,\qquad\qquad \bar\ell_4
  \equiv\log\left(\frac{\Lambda_4^2}{m_{\pi}^2}\right) = 4.61(4)(11)\, ,
\end{equation}
where we set $m_\pi=139.6\ \mathrm{MeV}$ conventionally. The first
error is statistical, the second systematical (see
Ref.~\cite{roberto} for details on how we estimate the systematic
errors). The statistical accuracy on these estimates for
$\bar{\ell}_{3,4}$ is quite impressive and the mean values are in good
agreement with the literature (see
Refs.~\cite{Leutwyler:2006qq,silvia} for recent reviews).

We conclude by mentioning that the values of $\bar{\ell}_{3,4}$
determined from the aforementioned $\chi$PT fit for $f_\mathrm{PS}$
and $m_\mathrm{PS}$ in the continuum~\cite{roberto} are in good
agreement with the values quoted in Eq.~(\ref{eq:lbars}).

\subsection{Nucleon Mass}

\begin{figure}[t]
  \centering
  \subfigure[\label{fig:nucleon}]%
  {\includegraphics[width=0.415\linewidth]{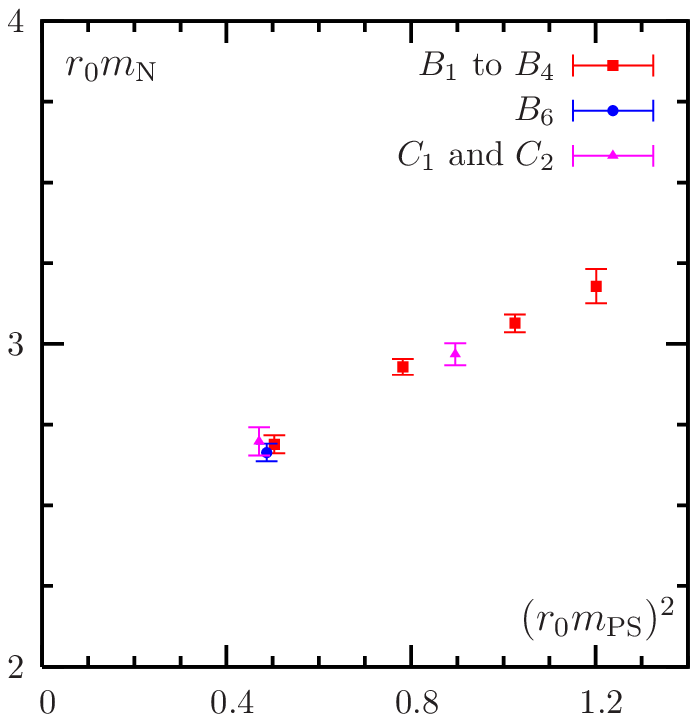}}
  \quad
  \subfigure[\label{fig:splitting}]%
  {\includegraphics[width=0.45\linewidth]{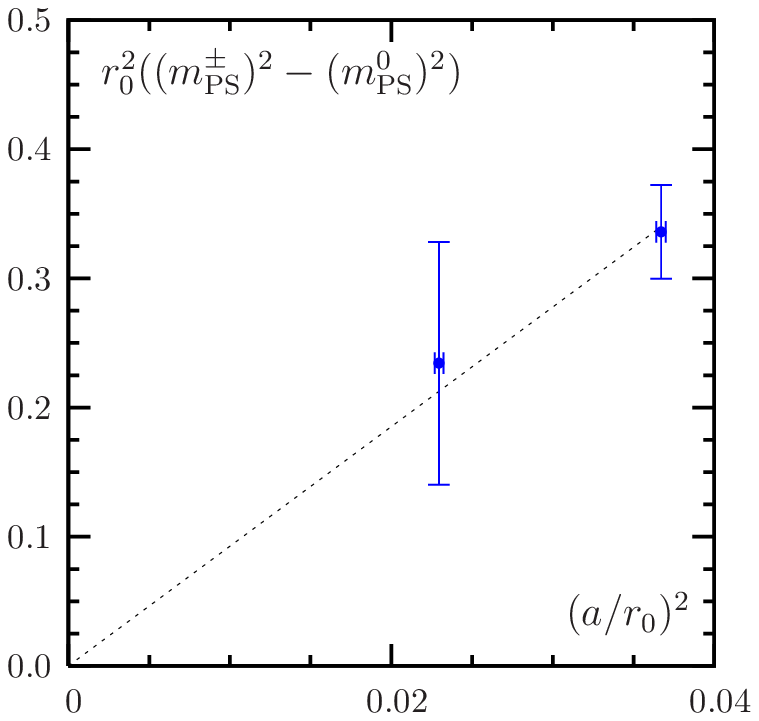}}
  \caption{(a) Nucleon mass $r_0m_\mathrm{N}$ as a function of
    $(r_0m_\mathrm{PS})^2$ for $\beta=3.9$ and $\beta=4.05$.
    (b) $(r_0^2((m_\mathrm{PS}^\pm)^2-(m_\mathrm{PS}^0)^2)$ as function
  of $(a/r_0)^2$ for ensembles $B_1$ and $C_1$.}
  \label{fig:d}
\end{figure}

In figure~\ref{fig:nucleon} we plot our preliminary results for the
nucleon mass $m_\mathrm{N}$ in units of $r_0$ as a function of
$(r_0m_\mathrm{PS})^2$. Also in this quantity scaling violations are
not visible within the statistical accuracy. Moreover, as shown by the
comparison of ensemble $B_1$ and $B_6$, finite size effects seem to be
negligible. Concentrating on $\beta=3.9$ only, since we do not yet have enough
data points at $\beta=4.05$, we fit the leading one-loop Heavy Baryon
$\chi$PT formula~\cite{Jenkins:1990jv,Becher:1999he}
\begin{equation}
  \label{eq:nucleon}
  m_\mathrm{N} = m_\mathrm{N}^0 + 4c_1m_\mathrm{PS}^2 -
  \frac{3g_A^2}{32\pi f_\mathrm{PS}^2}m_\mathrm{PS}^3
\end{equation}
with $g_A=1.267$ and the two free parameters $m_\mathrm{N}^0$ and
$c_1$ to our data. The fit provides a good description of the data with
$\chi^2/\mathrm{dof}=0.2$. With the physical nucleon mass as input the
value of the lattice spacing at $\beta=3.9$ can be determined to
$a=0.0879(12)\ \mathrm{fm}$ in very good agreement with the
determination from $f_\pi$. This result not only successfully cross
checks the determination of the lattice spacing in physical units from
$f_\pi$. If confirmed, it also represents a test of QCD as the theory
of strong interactions. For more details, including the error
determination and other baryon masses see Ref.~\cite{dina}.

\subsection{Flavour Symmetry Breaking Effects}

As mentioned in section \ref{sec:action}, flavour symmetry is
explicitly broken. As a consequence there is a potential difference of
$\mathcal{O}(a^2)$ between the masses of charged and neutral pseudo 
scalar mesons. Notice that to the latter also disconnected diagrams
contribute. We have determined the mass of the neutral pseudo scalar
meson for the ensembles $B_1$, $B_3$, $B_6$, $C_1$ and $C_2$. 
The results are shown in table~\ref{tab:mpi0} where we also report the
the corresponding values for the charged meson mass. 

The neutral pseudo scalar meson is lighter than the charged one. This
observation is consistent with $\chi$PT predictions for the observed
first order phase transition scenario (see Ref.~\cite{Shindler:2007vp}
and references therein). For ensembles $B_1$ and $C_1$ we plot in
figure~\ref{fig:splitting} the difference 
\begin{equation}
  \label{eq:split}
  r_0^2((m_\mathrm{PS}^\pm)^2 -(m_\mathrm{PS}^0)^2)
\end{equation}
as a function of $(r_0/a)^2$. The quantity (\ref{eq:split}) is
expected to scale linearly in $(r_0/a)^2$ towards the continuum,
which is confirmed by our data. The dashed line in
figure~\ref{fig:splitting} is not a fit, but 
it is there only to guide the eye.

Even if this analysis provides evidence for the expected scaling
behaviour of the pion mass splitting, the effect still amounts to
about $16\%$ at our smallest value of the lattice spacing and charged
$m_\mathrm{PS}\sim300\ \mathrm{MeV}$. If compared to the ``natural''
size  one would expect for $\mathcal{O}(a^2)$ cut-off effects, which
is of the order of $a^2\Lambda_\mathrm{QCD}^4$, one has to conclude
that this effect is unexpectedly large and one would like to be able
to better understand it. Of particular interest is the question
whether this effect arises \emph{dynamically}, or is due to large 
coefficients in the Symanzik expansion. In the latter case one might
expect that this effect is not restricted to only the neutral pseudo
scalar meson mass.

Note that all other possible splittings we have determined so far are
negligible. For instance the splittings in the vector
and $\Delta$ channels appear to be consistent with zero (see
table~\ref{tab:mpi0} for $m_\mathrm{V}$ and Ref.~\cite{dina} for the
$\Delta$). Similarly, the 
difference between the decay constants of charged and neutral pseudo
scalar meson is negligible. 
 
Currently, the ETM collaboration is investigating this question
\cite{giancarlo,etmc:asqr}, and from a theoretical point of view, an
analysis \`a la Symanzik of the 
charged and the neutral pseudo scalar meson masses lead to the formulae 
\begin{equation}
  \label{eq:sym}
  \begin{split}
    (m_\mathrm{PS}^0)^2 &= m_\pi^2 + a^2\zeta_\pi +
    \mathcal{O}(a^2m_\pi^2,a^4)\\ 
    (m_\mathrm{PS}^\pm)^2 &= m_\pi^2 +
    \mathcal{O}(a^2m_\pi^2,a^4)\\ 
  \end{split}    
\end{equation}
which show that the difference
$(m_\mathrm{PS}^0)^2-(m_\mathrm{PS}^\pm)^2$  
is given by the term proportional to $\zeta_\pi$. Here
$\zeta_\pi\equiv\langle\pi^0|\mathcal{L}_6|\pi^0\rangle$ and
$\mathcal{L}_6$ is the dimension six term in the Symanzik 
effective Lagrangian.

\begin{table}[t!]
  \centering
  \begin{tabular*}{.8\textwidth}{@{\extracolsep{\fill}}lcccc}
    \hline\hline
    Ensemble & $am_\mathrm{PS}^\pm$ & $am_\mathrm{PS}^0$ &
    $am_\mathrm{V}^\pm$ & $am_\mathrm{V}^0$ \\
    \hline\hline
    $B_1$ & $0.13623(65)$ & $0.111(6)$ & $0.404(22)$ & $0.395(17)$ \\
    $B_3$ & $0.19403(50)$ & $0.167(9)$ & $0.428(08)$ & $0.419(17)$ \\
    $B_6$ & $0.13377(24)$ & $0.110(8)$ & $0.416(12)$ & $0.400(25)$ \\
    \hline
    $C_1$ & $0.1038(6)$ & $0.091(11)$ & $0.337(20)$ & $0.372(29)$ \\
    $C_2$ & $0.1432(6)$ & $0.126(06)$ & $0.337(12)$ & $0.346(12)$ \\
    \hline\hline
  \end{tabular*}
  \caption{Comparison of values for the charged and the neutral pseudo
  scalar (vector) meson masses.}
  \label{tab:mpi0}
\end{table}

The main result of the analysis of Refs.~\cite{giancarlo,etmc:asqr}
is that $\zeta_\pi$ is a large number which in the 
vacuum saturation approximation can be estimated 
to be proportional to $|\hat{G}_\pi|^2$, where 
$\hat{G}_\pi =\langle 0 |\hat{P}^3|\pi^0\rangle$. The latter matrix
element is numerically large: one finds
$|\hat{G}_\pi|^2/\Lambda_\mathrm{QCD}^4$ around
$20-25$~\cite{giancarlo,etmc:asqr}. This result can provide an
interesting physical
explanation for the large splitting observed in the pseudo scalar
masses. Moreover, since it can be shown that $\zeta_\pi$ enters only
the  neutral pseudo scalar mass (and related quantities), one also
finds a possible explanation of why all other splittings determined so
far turn out to be small. In addition it provides hope that in the
future we shall not find large lattice artifacts due to flavour breaking.

At this point the question might arise whether the large splitting in
the pseudo scalar masses affects the $\chi$PT fits and in particular
the $\chi$PT FS estimates. At the current level of our analysis, where
we work under the assumption that lattice artifacts are zero in
charged $m_\mathrm{PS}$ and $f_\mathrm{PS}$ -- and we have shown good 
theoretical and numerical evidence for this -- all the fits are
performed in the continuum, where flavour symmetry is restored. Hence,
we do not expect any effect of flavour symmetry breaking on our
analysis with $\chi$PT.

\subsection{Further ETMC Results}

Our current simulations contain two light quark flavours with
degenerate mass in the sea. In the unitary set-up we can hence
determine the average up-down quark mass in the
$\overline{\mathrm{MS}}$ scheme by using $\chi$PT fits discussed in
section \ref{subsec:fit}. The result is
$m_\mathrm{ud}^{\overline{\mathrm{MS}}}(\mu_R=2\
\mathrm{GeV})=3.62(10)(23)\ \mathrm{MeV}$.
We have again used the estimate for $Z_\mathrm{P}$ coming from RI-MOM
\cite{cecilia,etmc:rimom}. Here and in the following the first error
is statistical, the second systematical. 

In order to also determine
the strange and possibly the charm quark masses, we have to use a
partially quenched set-up where we compute propagators on the
available gauge configurations with 
several values of the valence quark mass -- which are now different
from the sea quark mass -- around the strange and the charm quark
mass. The set-up and the calculation are described in detail in
Refs.~\cite{etmc:light,cecilia} and the result for the strange quark
mass for $\beta=3.9$ only reads 
$m_{s}^{\overline{\textrm{MS}}}(\mu_R=2~{\rm GeV})=
105(3)(8)\ \mathrm{MeV}$.
A comparison to other lattice QCD determinations of $m_{s}$ shows that
non-perturbative renormalisation has significant impact on the final
result. Estimates obtained with perturbative renormalisation tend to
be significantly lower than estimates with non-perturbatively computed
renormalisation constants. For more details as well as for further
related results, such as the kaon decay constant $f_\mathrm{K}$ and
the ratio $f_\mathrm{K}/f_\pi$, we refer to
Ref.~\cite{etmc:light,cecilia}.

In Ref.~\cite{benoit} we present our current estimates for the
charm quark mass at $\beta=3.9$. Our result reads
$m_{c}^{\overline{\textrm{MS}}}(\mu_R=m_c^{\overline{\textrm{MS}}})
= 1.481(21)(94)\ \mathrm{GeV}$. In Ref.~\cite{benoit} also results for
$f_D$ and $f_{D_s}$ are presented, which are in good agreement with
experiment.

In addition, the gauge configurations at $\beta=3.9$ have been used to
calculate a variety of 3-point correlation functions relevant for
semileptonic weak decays and electromagnetic transitions of light and
heavy-light pseudo scalar mesons. Results for the the vector, scalar
and tensor form factors of the pion, the vector and scalar form
factors relevant for Kl3 decay and for the Isgur-Wise function in the
heavy-quark limit can be found in Ref.~\cite{silvano}. On the same set
of gauge configurations there are preliminary results for the first
moment of the pion quark distribution function available. They can 
be found in \cite{zhaofeng} together with an effective stochastic
method to determine this quantity in lattice simulations.

The ETM collaboration has made a serious attempt to determine
properties of flavour singlet mesons with Mtm-LQCD, a first account of
which can be found in Ref.~\cite{Michael:2007vn}. The most interesting result
of this investigation is that the mass of the $\eta_2$ meson (the one
related to the anomaly in two flavour QCD, i.e. not a Goldstone boson) is 
consistent with a constant behaviour in the chiral limit and the value
around $800\ \mathrm{MeV}$ is compatible with expectations from a
model computation \cite{McNeile:2000hf}. (See also Ref.~\cite{craig})

For certain quantities, like $B_K$, it is crucial that the lattice
formulation exhibits good chiral properties. The overlap operator,
which obeys exact chiral symmetry at finite value of the lattice
spacing, can be used in a mixed action approach as a valence operator
on a maximally twisted mass sea. An exploratory study and first
results profiting from the chiral properties in the valence sector can
be found in Ref.~\cite{luigi}. 

A systematic uncertainty we cannot control at the moment is due to the
fact that the effects of the strange quark are not taken into account
in the simulations. In order to include these effects in Mtm-LQCD,
maintaining $\mathcal{O}(a)$ improvement at the same time, a $1+1$
split heavy doublet of quarks has to be simulated in addition 
to the mass degenerate light quark doublet \cite{Frezzotti:2003xj}. In
an exploratory study, published in Ref.~\cite{Chiarappa:2006ae}, it
was shown that this 
approach is feasible and that tuning is possible. However, it was also
shown that the effect of the first order phase transition, mentioned
in previous sections, strengthens significantly when the heavy doublet
is added. In Ref.~\cite{Jansen:2007sr} we report on an attempt to cure this
potential problem by using stout smearing~\cite{Morningstar:2003gk}. 
Our preliminary results suggest that stout smearing reduces indeed the
effects of the phase transition significantly.

An interesting investigation at tree level of perturbation theory is
presented in Ref.~\cite{jen}. In this framework cut-off effects can be
studied using analytic calculations. Scaling properties of Wilson and
Wilson twisted mass fermions are compared and it is shown
that maximally twisted mass fermions scale with a rate of 
$\mathcal{O}(a^2)$.

For a study using Mtm-LQCD for simulations of QCD thermodynamics
see Ref.~\cite{Ilgenfritz:2007qr}.

\subsection{Exploring the $\epsilon$-regime with Maximally Twisted Mass Fermions}

A slightly different direction as compared to all the aforementioned
results is explored by performing studies of the $\epsilon$-regime
\cite{Gasser:1987ah} with Mtm-LQCD at $\beta=3.9$
\cite{andrea}. Simulations in the $\epsilon$-regime are not restricted
to formulations with exact chiral symmetry. And, since the 
twisted mass parameter provides -- as discussed before -- an infra-red
cut-off to the eigenvalue spectrum of the twisted mass Dirac operator
there is also no technical complication to expect, unlike that arises
e.g. for Wilson fermions when the quark mass becomes too small. 

Preliminary results of this investigation presented in
Ref.~\cite{andrea} are quite encouraging. Simulations turn out to be
feasible, perform smoothly and are much less computer time demanding
than simulations with e.g. the overlap operator. A first result of
this study at $\beta=3.9$ is an estimate for the chiral condensate
$\Sigma^{\overline{\mathrm{MS}}}(\mu_R=2\
\mathrm{GeV})=(264(12)(4)^{+20}_{-0}\ \mathrm{MeV})^3$, which is in
perfect agreement with the value determined from the $\chi$PT fit
$\Sigma^{\overline{\mathrm{MS}}}(\mu_R=2\
\mathrm{GeV})=(267(4)(7)\ \mathrm{MeV})^3$.

\section{Summary}

In this proceeding contribution we have summarised the current status
of the two flavour project of the European Twisted Mass
collaboration. The collaboration has generated gauge configurations
using a doublet of mass degenerate Wilson twisted mass fermions at
maximal twist for three different values of the lattice spacing
$a\lesssim0.1\ \mathrm{fm}$, volumes with physical extent larger than
$2\ \mathrm{fm}$ and values for the (charged) pseudo scalar meson mass
in the range of $300$ to $700\ \mathrm{MeV}$. The results for the two
finer lattice spacings of about $a\sim0.086\ \mathrm{fm}$ and
$a\sim0.067\ \mathrm{fm}$ are close to final, while the results for the
coarsest lattice spacing are still in a preliminary state.

We have presented results for (charged) $f_\mathrm{PS}$ and
$m_\mathrm{PS}$ and a 
scaling analysis for $f_\mathrm{PS}$ at fixed, but finite volume
indicating that $\mathcal{O}(a)$ improvement in Mtm-LQCD works very
well. Lattice artifacts turn out to be compatible with zero to our
current statistical accuracy, which is of the order of $1\%$.

We have shown evidence that finite size effects in $f_\mathrm{PS}$ and
$m_\mathrm{PS}$ can be described by means of formulae derived in
chiral perturbation theory to the level of statistical accuracy of the
data. It turns out that the asymptotic L{\"u}scher formula presented
in Ref.~\cite{Colangelo:2005gd} works better than the NLO formula from
Ref.~\cite{Gasser:1986vb}. 

NLO continuum chiral perturbation theory \cite{Gasser:1986vb} can be
used successfully to describe the quark mass dependence of mass and
decay constant of the charged pseudo scalar meson. The low energy
constants $B_0$, $F_0$, $\bar{\ell}_3$ and $\bar{\ell}_4$ can be
determined to high statistical accuracy. The corresponding fit can
also be used to determine the lattice spacings using the physical
values of $f_\pi$ and $m_\pi$. The results are in good agreement with
the determination from the nucleon mass.

There is theoretical and numerical evidence that large flavour
breaking effects appear only in the mass of the neutral pseudo scalar
meson and trivially related quantities. In particular, all other
flavour splittings measured so far turn out to be compatible with
zero. In addition the mass splitting in the pseudo scalar meson masses
scales as expected towards the continuum limit.

The collaboration is analysing the available gauge configurations for
many more physical quantities. As an example we have presented first
results for the strange and the charm quark masses. For the strange
quark mass, where we can compare to other lattice determinations, it
turns out that the difference between perturbative and
non-perturbative renormalisation is significant. Quantities like
$f_\mathrm{K}/f_\pi$ or $f_\mathrm{D}/f_{\mathrm{D}_s}$ are in good
agreement with experiment.

In the future we plan to repeat all these calculations with strange and
charm quark effects taken into account in the simulations. The ETM
collaboration is currently investigating the optimal set-up 
for simulations with $2+1+1$ dynamical quark flavours and maximal
twist. Algorithms and codes are available and the simulations are due
to start.

We conclude by mentioning that all ETMC ensembles are stored on
ILDG disk space~\cite{detar}. They are available to non-ETMC
members on a request basis, whenever there is no overlap to ongoing
ETMC projects.

\section*{Acknowledgements}

I would like to thank all members of ETMC for the most enjoyable
collaboration and for making all the results available to me. I would
like to thank the LOC of Lattice 2007 for 
organising a very nice conference and for giving  me the opportunity
to give this talk.  Special thanks to K.~Jansen and C.~Michael for
their steady support, help, encouragement  and many discussions. When
preparing the talk and this proceedings I enjoyed many useful and
interesting discussions with C.~Alexandrou, P.~Dimopoulos,
R.~Frezzotti, G.~Herdoiza, V.~Lubicz, C.~McNeile, G.~M{\"u}nster,
G.C.~Rossi, L.~Scorzato, A.~Shindler, M.~Wagner and U.~Wenger. I am
indebted to the ILDG working groups and in particular B.~Orth,
D.~Pleiter, H.~St{\"u}ben and S.~Wollny for their support. This work
has been supported by the EU Integrated Infrastructure Initiative
Hadron Physics (I3HP) under contract RII3-CT-2004-506078.

\bibliographystyle{h-physrev4}
\bibliography{bibliography}

\end{document}